# QUASI-OPTIMAL FILTERING IN INVERSE PROBLEMS

V. YU. TEREBIZH*

*Sternberg Astronomical Institute, Universitetskij Prospekt 13, Moscow 119992, Russia*

A way of constructing a nonlinear filter close to the optimal Kolmogorov–Wiener filter is proposed within the framework of the statistical approach to inverse problems. Quasi-optimal filtering, which has no Bayesian assumptions, produces stable and efficient solutions by relying solely on the internal resources of the inverse theory. The exact representation is given of the feasible region for inverse solutions that follows from the statistical consideration.

*Keywords*: Inverse problems; Image restoration

## 1 INTRODUCTION

Many problems of physics and engineering, and in particular the image restoration problem, proceed from a linear model

$$y_0 = \mathbf{H}x_0 + \boldsymbol{\xi},$$
$$\langle \boldsymbol{\xi} \rangle = \boldsymbol{a}, \quad \text{cov}(\boldsymbol{\xi}) \equiv \langle (\boldsymbol{\xi} - \boldsymbol{a})(\boldsymbol{\xi} - \boldsymbol{a})^{\text{T}} \rangle = \mathbf{C}, \tag{1}$$

where the $n \times 1$ vector $x_0$ is an unknown object, which has to be restored, the $m \times n$ matrix $\mathbf{H}$ is the *point spread function* (PSF), the $m \times 1$ vector $y_0$ is the observed image, and $\boldsymbol{\xi}$ is the random noise with average level $\boldsymbol{a}$ and covariance $m \times m$ matrix $\mathbf{C}$. We assume that $m \geq n$; the angular brackets mean averaging on the probabilistic ensemble.

Both the maximum entropy and, in essence, the regularization methods of solving Eq. (1) are based on the Bayesian approach, which assigns a prior probability to possible inverse solutions (Janes, 1957a, b; Phillips, 1962; Tikhonov, 1963; Tikhonov and Arsenin, 1977; Press *et al.*, 1992; Jansson, 1997). In the statistical approach to inverse problems, the sought solution $\tilde{x}$ is considered as a statistical estimate of the deterministic object $x_0$, given its image, a PSF, properties of the noise and available *a priori* information about the object (Terebizh, 1995a, b). Being a function of the stochastic image $y_0$, the inverse solution $\tilde{x}$ is also a random vector, as a rule, with mutually dependent components. We adhere below to the second of these points of view.

To define properly the notion of the quality of an estimate, we should carry out two preliminary procedures: Firstly, correlations in the image $y_0$ should be eliminated; secondly, the dimension of the vector that describes the misfit with the observed image to the object's length $n$ should be reduced.

---

* Present address: Nauchny 98409, Crimea, Ukraine. E-mail: terebizh@crao.crimea.ua

The first of the procedures is based on the known linear transform

$$z_0 = \mathbf{C}^{-1/2}(y_0 - a), \quad \eta = \mathbf{C}^{-1/2}(\xi - a), \quad \mathbf{A} = \mathbf{C}^{-1/2}\mathbf{H}, \qquad (2)$$

which converts the general model (1) to the standard model

$$z_0 = \mathbf{A}x_0 + \eta,$$
$$\langle \eta \rangle = 0, \quad \text{cov}(\eta) = \mathbf{E}_m, \qquad (3)$$

where $\mathbf{E}_m$ is the unit $m \times m$ matrix. The matrix $\mathbf{C}^{-1/2}$ in Eq. (2) is inverse to the square root of $\mathbf{C}$, which implies a positive spectrum of the covariance matrix.

The second step proceeds from the singular value decomposition of the matrix $\mathbf{A}$ (see for example Golub and Van Loan (1989) and Press *et al.* (1992)). Assume that rank $(\mathbf{A}) = n$. Then

$$\mathbf{A} = \mathbf{U}\mathbf{\Delta}\mathbf{V}^{\mathrm{T}}, \qquad (4)$$

where $\mathbf{U}$ is an $m \times n$ column-orthogonal matrix, $\mathbf{\Delta}$ is a diagonal $n \times n$ matrix with positive singular values $\boldsymbol{\delta} = [\delta_1, \ldots, \delta_n]^{\mathrm{T}}$ of $\mathbf{A}$, placed in the order of their decrease, and an $n \times n$ matrix $\mathbf{V} = [v_1, \ldots, v_n]$ is orthogonal:

$$\mathbf{U}^{\mathrm{T}}\mathbf{U} = \mathbf{E}_n, \quad \mathbf{\Delta} = \text{diag}(\boldsymbol{\delta}), \quad \mathbf{V}^{-1} = \mathbf{V}^{\mathrm{T}}. \qquad (5)$$

The corresponding decomposition of the object $x_0$ in the eigenvectors system $\{v_k\}$, namely

$$x_0 = \mathbf{V}p_0 = \sum_{k=1}^{n} p_{0k}v_k, \quad p_0 = \mathbf{V}^{\mathrm{T}}x_0, \qquad (6)$$

defines the vector $p_0$ of the object's principal components. Like the familiar Fourier coefficients, the principal components are often easier to recover than the object itself. The multiplication if Eq. (3) by $\mathbf{U}^{\mathrm{T}}$ is similar to the application of the Fourier transform. Designating

$$\boldsymbol{\phi} = \mathbf{U}^{\mathrm{T}}z_0, \quad \boldsymbol{\zeta} \equiv \mathbf{U}^{\mathrm{T}}\eta, \qquad (7)$$

we obtain a final representation of the linear model:

$$\boldsymbol{\phi} = \mathbf{\Delta}p_0 + \boldsymbol{\zeta},$$
$$\langle \boldsymbol{\zeta} \rangle = 0, \quad \text{cov}(\boldsymbol{\zeta}) = \mathbf{E}_n. \qquad (8)$$

The advantages of use of the 'refined image' $\boldsymbol{\phi}$ of length $n$ are especially appreciable when $m \gg n$.

## 2 FEASIBLE REGION

Assume, for simplicity, that the noise $\boldsymbol{\xi}$ is a Gaussian deviate. Then $\{\zeta_k\}$ in Eq. (8) are independent Gaussian deviates with zero mean value and unit variance, and the random variable $\|\boldsymbol{\phi} - \mathbf{\Delta}p_0\|^2 = \sum_{k=1}^{n} \zeta_k^2$ has a $\chi^2$ distribution with $n$ degrees of freedom (Cramer, 1946,

Chapter 18). This result allows us to introduce a similar random variable, namely the misfit

$$\Theta(y_0|x) \equiv \|\boldsymbol{\phi} - \boldsymbol{\Delta p}\|^2, \tag{9}$$

as a measure of the quality of a trial object's estimate $x = \mathbf{V}p$.

Let $t_\gamma^{(n)} \geq 0$ be a quantile of the $\chi_n^2$ distribution $P_n(t)$, that is the root of equation $P_n(t) = \gamma$. Just as is usually done in mathematical statistics (Cramer, 1946), we shall choose the appropriate boundary significance levels for an inverse solution $\alpha_1$ and $\alpha_2$ ($0 \leq \alpha_1 \leq \alpha_2 \leq 1$). By definition, a trial object's estimate $x$ is called feasible, if

$$t_{1-\alpha_2}^{(n)} \leq \Theta(y_0|x) \leq t_{1-\alpha_1}^{(n)}. \tag{10}$$

We simply require of a feasible estimate $x$ that its image $y(x)$ should have moderate deviation, in the statistical sense, from the observed image $y_0$. Inequalities (10) define the *feasible region* (FR), consisting of all the object's estimates $\{x\}$ that have feasible agreement with the data. It is convenient to call $x$ the estimate of a significance level $\alpha$, if the misfit $\Theta(y_0|x) = t_{1-\alpha}^{(n)}$, that is

$$\|\boldsymbol{\phi} - \boldsymbol{\Delta p}\|^2 = t_{1-\alpha}^{(n)}. \tag{11}$$

According to the well-known Gauss–Markov theorem, the *least-squares estimate* (LSE)

$$x_* = (\mathbf{A}^T\mathbf{A})^{-1}\mathbf{A}^T z_0 \tag{12}$$

has the smallest variance of all the unbiased object's estimates (Lawson and Hanson, 1974). It follows from Eqs. (4) and (12) that

$$x_* = \mathbf{V}p_* = \sum_{k=1}^n p_{*k} v_k, \quad p_* = \boldsymbol{\Delta}^{-1}\boldsymbol{\phi}. \tag{13}$$

Equation (13) define the principal components $p_*$ of LSE. Unlike the object principal components $\{p_{0k}\}$, the LSE components $\{p_{*k}\}$ are random variables. One can easily find the mean value and the covariance matrix of the LSE:

$$\langle p_* \rangle = p_0, \quad \text{cov}(p_*) = \boldsymbol{\Lambda}^{-1}, \tag{14}$$

where the matrix

$$\boldsymbol{\Lambda} \equiv \boldsymbol{\Delta}^2 = \text{diag}(\lambda_1, \ldots, \lambda_n), \quad \lambda_k = \delta_k^2. \tag{15}$$

Thus, the LSE principal components $\{p_{*k}\}$ are the unbiased estimates of $p_{0k}$, and $\text{var}(p_{*k}) = \lambda_k^{-1}$. Usually, the 'tail' of the sequence $\{\lambda_k\}$ is very small; so the variance of corresponding $\{p_{*k}\}$ and consequently the variance of LSE are huge.

Let us recall the geometrical interpretation of this phenomenon. With the help of Eqs. (4) and (13), it is easy to transform definition (11) into the form

$$(x - x_*)^T \mathbf{I}(x - x_*) = t_{1-\alpha}^{(n)}, \quad \mathbf{I} = \mathbf{A}^T\mathbf{A} = \mathbf{V}\boldsymbol{\Lambda}\mathbf{V}^T. \tag{16}$$

Therefore, the FR consists of hollow ellipsoids, centred at the LSE, and the shape of ellipsoids is defined by the $n \times n$ Fisher matrix $\mathbf{I}$ (Terebizh, 1995a, b). Small values of the farthest eigenvalues $\{\lambda_k\}$ in the spectrum of matrix **I** give rise to an extremely elongated shape for the FR. Just that phenomenon reveals itself in the well-known instability of inverse solutions.

The FR usually does not include the LSE and the manifold in its vicinity. The reason is that the object's estimates close to LSE try to 'explain' all details of the observed image, irrespective of their statistical significance. Since the model (1) supposes essential smoothing of the object, one should admit large erroneous oscillations in the object's estimate in order to fit tiny random fluctuations in the image.

## 3  OPTIMAL LINEAR FILTER

It is possible to mitigate the harmful influence of the small eigenvalues of the Fisher matrix by introducing into Eq. (13) the appropriate set of weights $\boldsymbol{w} = [w_1, \ldots, w_n]^{\mathrm{T}}$; so

$$\boldsymbol{x}_w \equiv \sum_{k=1}^{n} w_k p_{*k} \boldsymbol{v}_k = \mathbf{V}\mathbf{W}\boldsymbol{p}_*, \quad \mathbf{W} = \mathrm{diag}(\boldsymbol{w}). \tag{17}$$

A number of known inverse solutions, in particular, the optimal estimate given by Kolmogorov (1941) and Wiener (1942), the regularized solution given by Phillips (1962) and Tikhonov (1963), and the truncated estimate given by Varah (1973), Hansen (1987, 1993) and Press *et al.* (1992), belong to the class of linearly filtered estimates. It follows from Eqs. (6) and (17) that the squared error of the filtered estimate is

$$\varepsilon_w^2 \equiv \langle \|\boldsymbol{x}_w - \boldsymbol{x}_0\|^2 \rangle = \sum_{k=1}^{n} \left( \frac{w_k^2}{\lambda_k} + (1 - w_k)^2 p_{0k}^2 \right). \tag{18}$$

As one can see, the error is minimized by the set of weights

$$\tilde{w}_k = \frac{\lambda_k p_{0k}^2}{1 + \lambda_k p_{0k}^2}, \quad k = 1, 2, \ldots, n, \tag{19}$$

which constitutes the optimal Wiener filter $\tilde{\mathbf{W}}(\boldsymbol{p}_0) = \mathrm{diag}[\tilde{\boldsymbol{w}}(\boldsymbol{p}_0)]$. Consequently, the best of linearly filtered estimates of the object is

$$\tilde{\boldsymbol{x}}_w = \sum_{k=1}^{n} \tilde{w}_k p_{*k} \boldsymbol{v}_k = \mathbf{V}\tilde{\boldsymbol{p}}_w, \quad \tilde{\boldsymbol{p}}_w = \tilde{\mathbf{W}}(\boldsymbol{p}_0)\boldsymbol{p}_*. \tag{20}$$

An important feature of the optimal filter is that the weights $\tilde{\boldsymbol{w}}$ depend not only on the known properties of the PSF and the noise but also upon the object itself. For that reason, the filter can be applied only in the Bayesian approach to inverse problems. It is worth noting, in this connection, that the investigations of Kolmogorov (1941) and Wiener (1942) focused on time series analysis, where the Bayesian approach is well justified since the Gaussian nature of ensembles is ensured by the central limit theorem. For most other inverse problems, and in particular, image restoration, the availability of both object ensembles and prior probability distributions on those ensembles is unnatural.

We can simplify the general description of the FR for the linearly filtered estimates by substituting $\boldsymbol{p}_w = \mathbf{W}\boldsymbol{p}_*$ in Eq. (11) or in Eq. (16). The result is

$$\|[\mathbf{W} - \mathbf{E}_n]\boldsymbol{\phi}\|^2 = t_{1-\alpha}^{(n)}. \tag{21}$$

This condition imposes restrictions on the system of weights $w$. Then Eq. (17) enables the filtered estimate to be found.

One can expect that the requirement (21) is satisfied for the optimal filter $\mathbf{W} = \tilde{\mathbf{W}}(p_0)$ at moderate values of the significance level $\alpha$. Indeed, extensive numerical simulations are in agreement with this assumption; the corresponding significance level usually is more than 0.70.

## 4 QUASI-OPTIMAL FILTER

If it were possible to find a good approximation of the object's principal components $\{p_{0k}\}$ in Eq. (19) with only the given and the observed quantities, the corresponding filter would doubtless have a practical value, but we have no *a priori* information for such immediate approximation. At the same time, and that is the key point of the quasi-optimal filtering, we have enough information about the structure of the optimal estimate $\tilde{x}_w$, in order to require similar properties for the estimate of the object searched for.

By substituting $x_0$ from Eq. (6) and $\tilde{x}_w$ from Eq. (20) into Eq. (18), and noting that the orthogonal transform does not change the vector norm, we obtain

$$\langle \|\tilde{\mathbf{W}}(p_0)p_* - p_0\|^2 \rangle = \tilde{\varepsilon}_w^2(p_0). \tag{22}$$

This equation simply gives another representation of the error of the optimal filter, which, by definition, is the smallest in the class of linear filters.

Let us now consider a trial estimate $p$ close to $p_0$ (Fig. 1). Taking into account Eq. (22), we shall require that the filter

$$\check{\mathbf{W}}(p) = \text{diag}[\check{w}(p)], \quad \check{w}_k(p) = \frac{\lambda_k p_k^2}{1 + \lambda_k p_k^2}, \tag{23}$$

which is based on such an estimate, had the minimal error:

$$\langle \|\check{\mathbf{W}}(p)p_* - p\|^2 \rangle = \min. \tag{24}$$

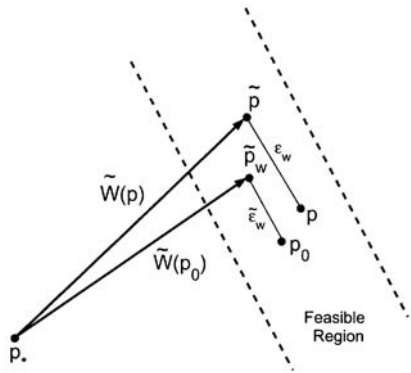

FIGURE 1  Schematic representation of the optimal and the quasi-optimal filtering in the space of principal components: $p_0$, object; $p_*$, LSE; $\tilde{\mathbf{W}}(p_0)$, optimal filter; $\tilde{p}_w$, optimal estimate of the object; $p$, trial estimate; $\check{\mathbf{W}}(p)$, Wiener filter for the trial estimate; $\check{p}$, quasi-optimal estimate of the object. The errors of the filters are shown by the segments $\tilde{\varepsilon}_w(p_0)$ and $\varepsilon_w(p)$.

Note that the quasi-optimal filter (23) has the same structure as the optimal Wiener filter (19). Thus, we search for the estimate that most closely simulates the behaviour of the best inverse solution.

If we depart from the averaging procedure, which is executable only in theory, and add the condition (21), which requires that the trial object's estimate belongs to the FR, we obtain the simultaneous conditions

$$\|[\tilde{\mathbf{W}}(\boldsymbol{p}) - \mathbf{E}_n]\boldsymbol{\phi}\|^2 = t_{1-\alpha}^{(n)},$$
$$\|\tilde{\mathbf{W}}(\boldsymbol{p})\boldsymbol{p}_* - \boldsymbol{p}\|^2 = \min. \quad (25)$$

The solution $\boldsymbol{p}_{\min}$ of this system allows us to find the quasi-optimal estimates of the object and its principal components:

$$\tilde{\boldsymbol{p}} = \tilde{\mathbf{W}}(\boldsymbol{p}_{\min})\boldsymbol{p}_*, \quad \tilde{\boldsymbol{x}} = \mathbf{V}\tilde{\boldsymbol{p}}. \quad (26)$$

Indeed, we are ultimately interested not in the $\boldsymbol{p}_{\min}$ that is intended to replace $\boldsymbol{p}_0$ only in argument of the filter (see Fig. 1), but in the filtered estimate of the principal components $\tilde{\boldsymbol{p}}$, which is analogous to the optimal Wiener estimate $\tilde{\boldsymbol{p}}_w$ in Eq. (20).

In the components of the corresponding vectors, Eq. (25) can be written as

$$\sum_{k=1}^{n} [\tilde{w}_k(\boldsymbol{p}) - 1]^2 \phi_k^2 = t_{1-\alpha}^{(n)},$$
$$\sum_{k=1}^{n} [\tilde{w}_k(\boldsymbol{p}) p_{*k} - p_k]^2 = \min, \quad (25')$$

where the $\tilde{w}_k(\boldsymbol{p})$ are given by Eq. (23), and $\phi_k$ are the components of the vector $\boldsymbol{\phi}$ defined by Eq. (7).

Unlike the Wiener filter, the quasi-optimal filter is nonlinear with respect to the LSE $\boldsymbol{p}_*$, because a solution $\boldsymbol{p}_{\min}$ of the system (25) is dependent upon $\boldsymbol{p}_*$, and then we should apply filtering according to Eq. (26).

Since both functionals in Eq. (25) are positive definite, and the second functional is non-degenerate, the solution of the constrained minimization problem (25) is unique (Press *et al*., 1992, Section 18.4).

To understand the sense of quasi-optimal filtering better, it is useful to bear in mind the following. The object and its LSE were held fixed in creating the optimal filter, and the filter structure has been optimized. On the contrary, Eqs. (23) and (24) fix the previously determined structure of the filter (and the LSE, of course), concentrating attention on the search for an appropriate estimate of the object. Such an approach seems to be quite justified because simultaneous searches for both the best filter and a good inverse solution are possible only if complete information about the object is available. The efficiency of the optimal filtering should be high enough in the vicinity of the unknown object; so we do have reason to fix a form of the best filter for an estimate close to the object.

## 5 MODEL CASES

Equations (23), (25) and (26) form the basis for an algorithm that can be programmed with a high-level programming language.[1] We deliberately consider here simple examples, in order to show distinction between two filters under discussion more clearly.

---

[1] The sample MatLab program is accessible on request.

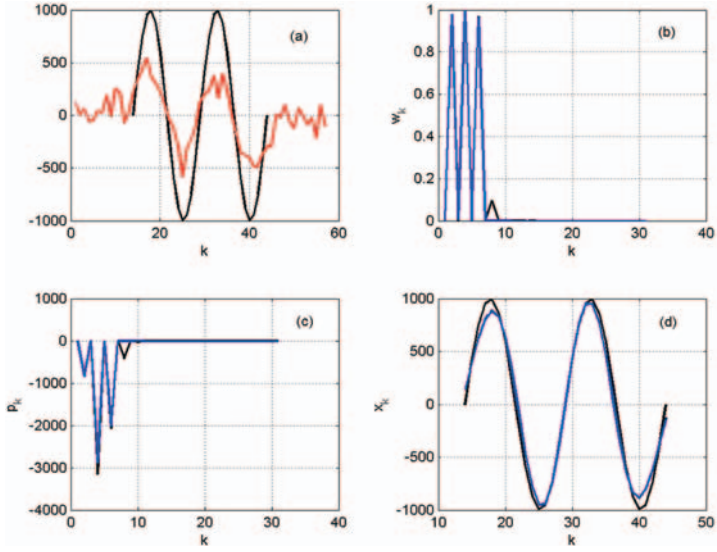

**FIGURE 2** (a) The object (black curve), and its blurred image (red curve). (b) Weights of the optimal Wiener (black curve) and the quasi-optimal (blue curve) filters. (c) Principal components of the object (black curve), and the quasi-optimal estimate (blue curve). (d) The object (black solid curve), the optimal (black dashed curve) and the quasi-optimal estimates (blue curve).

Figure 2 describes restoration of a low-frequency object that we have assumed to be the portion of a sinusoid having an amplitude of 1000. A space-invariant PSF

$$h(t - t') = R^{-1} \operatorname{sinc}^2 \left( \frac{t - t'}{R} \right) \tag{27}$$

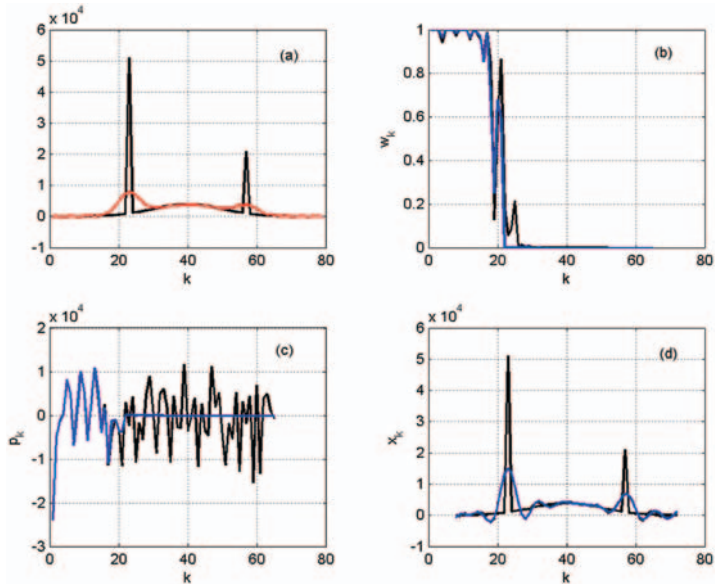

**FIGURE 3** (a) The object (black curve), and its blurred image (red curve). (b) Weights of the optimal Wiener (black curve) and the quasi-optimal (blue curve) filters. (c) Principal components of the object (black curve), and the quasi-optimal estimate (blue curve). (d) The object (black solid curve), the optimal estimate (blue dashed curve) and the quasi-optimal estimate (blue curve).

was adopted, where $\text{sinc}(t) \equiv [\sin(\pi t)]/\pi t$, and the characteristic radius $R$ was taken as 9 pixels. Function (27) can be considered as a one-dimensional analogue of the Airy diffraction pattern. The mean level of the Gaussian white noise $a$ was taken as zero, and its standard deviation $\sigma_\xi$ as 100. The significance levels of the filters were equal to each other.

Note the removal of the erroneous high-frequency oscillations in the object estimates, and the non-monotonic behaviour of both the optimal and the quasi-optimal weights, which is distinct from those for a truncated estimate. The quasi-optimal filter leaves in the object's estimate only those principal components that have the highest accuracy of restoration. The errors of the discussed filters were nearly the same for the considered example.

Figure 3 depicts a traditionally difficult model case that incorporates superposition of the sharp and smooth details. The Gaussian PSF has been applied this time with the standard deviation $\sigma_{\text{PSF}} = 3$ pixels; the noise standard deviation remained as above. As one can see from Figure 3, both the optimal and the quasi-optimal estimates have similar qualities.

The discussion of the non-negativity condition and the Poisson model has been given elsewhere (Terebizh, 2003).

## 6  CONCLUDING REMARKS

It is appropriate to emphasize importance of the Fisher matrix **I**, which plays a fundamental role not only in the linear model but also in the general inverse problem (Terebizh, 1995a, b). To simplify discussion, it was assumed above that the spectrum of the matrix **I** can include arbitrarily small but strictly non-zero eigenvalues $\{\lambda_k\}$. This restriction is not essential for the final results; the case of some zero eigenvalues can be treated with the aid of known additional procedures with the LSE (Press *et al.*, 1992).

From the viewpoint of the regularization theory, it might seem that the functional

$$F(\boldsymbol{p}) \equiv \|\tilde{\mathbf{W}}(\boldsymbol{p})\boldsymbol{p}_* - \boldsymbol{p}\|^2, \tag{28}$$

which we use in Eq. (25), is a stabilizing (smoothing) functional similar to $\|\boldsymbol{x}\|^2$ or to one of the several forms of the 'entropy' $\mathcal{E}(\boldsymbol{x})$. Indeed, the condition $F(\boldsymbol{p}) = \min$ promotes stabilization of the inverse solution, but the origin of this functional is of vital importance. The Bayesian approach proposes to compensate lack of *a priori* information by some general principle that directly concerns the properties of the sought object **x** itself. Obviously, it is possible to offer an unlimited number of such principles. We rely on the intrinsic reserves of the inverse theory. It appears that, instead of prior information on the object, it is enough to use a much weaker assumption that the optimal Wiener filter retains a high efficiency in the local vicinity of the unknown object.

*Acknowledgements*

The author is grateful to V. V. Biryukov (Moscow University), P. A. Jansson (University of Arizona), I. S. Savanov (Astrophysical Institute Potsdam) and A. A. Tokovinin (Cerro Tololo Inter-American Observatory) for valuable discussions.

*References*

Cramér, H. (1946) *Mathematical Methods of Statistics*, Princeton University Press, Princeton, New Jersey.
Golub, G. H., and Van Loan, C. F. (1989) *Matrix Computations*, Johns Hopkins University Press, Baltimore, Maryland.
Hansen, P. C. (1987) *BIT, Nord. Tidskr. Information beh.* **27**, 543.

QUASI-OPTIMAL FILTERING IN INVERSE PROBLEMS                                93

Hansen, P. C. (1993) *Regularization Tools*, Danish Computing Center for Research and Education, Technical University of Denmark, Copenhagen.

Jansson, P. A. (ed.) (1997) *Deconvolution of Images and Spectra*, Academic Press, San Diego, California.

Jaynes, E. T. (1957a) *Phys. Rev.* **106**, 620.

Jaynes, E. T. (1957b) *Phys. Rev.* **108**, 171.

Kolmogorov, A. N. (1941) *Proc. USSR Acad. Sci. (Math.)*, **5**, 3.

Lawson, C. L., and Hanson, R. J. (1974) *Solving Least Squares Problems*, Prentice-Hall, Englewood Cliffs, New Jersey.

Phillips, D. L. (1962) *J. Assoc. Comput. Mach.* **9**, 84.

Press, W. H., Teukolsky, S. A., Vetterling, W. T., and Flannery, B. P. (1992) *Numerical Recipes*, Cambridge University Press, Cambridge.

Terebizh, V. Yu. (1995a) *Phys. Usp.* **38**, 137.

Terebizh, V. Yu. (1995b) *Int. J. Imaging Syst. Technol.* **6**, 358.

Terebizh, V. Yu. (2003) *Bull. Crimean Astrophys. Obs.* **99**, 166.

Tikhonov, A. N. (1963) *Soviet Math. Dokl.* **4**, 1035.

Tikhonov, A. N., and Arsenin, V. Y. (1977) *Solutions of Ill-posed Problems*, Wiley, New York.

Varah, J. M. (1973) *SIAM J. Numer. Anal.* **10**, 257.

Wiener, N. (1942) NDRC Report, Massachusetts Institute of Technology, Cambridge, Massachusetts (Reprinted (1949) Wiley, New York).